# Photoelectric Effect at Sub-Photon Energy in Incident Pulsed Laser Radiation

**V. I. Kovalev[1,2]**

[1]P.N.Lebedev Physical Institute of the Russian Academy of Sciences, Leninsky pr. 53, Moscow, 119991, Russia,
[2]Heriot-Watt University, Edinburgh, EH14 4AS, UK
Tel.: +74991326262, fax: +74991357880. E-mail: kovalevvi@lebedev.ru

**Abstract:** Photoelectric effect in a *Ge*-on-*Si* single-photon avalanche detector (SPAD) at a sub-photon energy in incident pulsed laser radiation is considered in frames of classical electrodynamics of continuous media. It is shown that the energy of incident laser radiation, which is shared among a huge number of electrons in *Ge* matrix, can concentrate on only one of these through the effect of the constructive interference of the fields re-emitted by surrounding electrons. Conservation of energy in this case is upheld because of a substantial narrowing of the effective bandgap in heavily doped p-*Ge*, which is used in the design of considered SPAD.

______________________________________________________________________________

## 1. Introduction

Metrology of extremely low radiation energies/powers is the subject of vital importance for R&D in the area of Quantum Technologies. Single photon detectors (SPD) are the devices which do the job. Significant advances in both photoelectric semiconductor and thermal superconductor based SPDs have been achieved in recent years [1-3]. As conventionally was happening in the history of science and technologies, the more objects of study and more researches involved, the higher probability to find something new and unexpected. In the case of photoelectric SPDs this new is nonzero detection efficiency (DE) of such detectors at a sub-photon energy in the pulse of incident laser radiation [4,5]. Such observations are in direct contradiction with one of a few key effects, on which the quantum theory rests, and for which a classical description does not exist [6]. According to Einstein [7], a photoelectron can appear when the energy in light flux is less than one quantum or photon. The main postulates of his model for the photoelectric effect include 1) "light (*in the form of*) quanta are penetrating to the surface layer of a material", 2) "their energy ... is converted to the kinetic energy of electrons", and 3) "one light quantum gives up all its energy to one electron". This model, which was proven to be exceptionally productive since time of its appearance, is conventionally used as a justification for quantum nature of light and indivisibility of a quantum. Subsequent developments, however, reviled several features of the photoelectric effect, which were unaccountable in frames of the Einstein's model [8], and which stimulated further studies [9,10]. In this work a way is described to account for in frames of classical electrodynamics of continuous media the presence of photoelectric effect in *Ge*-on-*Si* single-photon avalanche diode (SPAD) [5] at a sub-photon energy in incident pulsed laser radiation.

## 2. Device design and characteristics

The structure of SPAD used in [5] is presented in Fig.1. It is presumed that an incident light, which enters the detector through high concentration boron-doped (~5×10$^{19}$ cm$^{-3}$) p++ *Ge* (p-*Ge*) layer of thickness $l = 0.1$ $\mu$m, is absorbed with creation of electron–hole pairs in 1 $\mu$m-thick layer of intrinsic *Ge* (i-*Ge*). The created PEs are then drugged by an applied voltage of ~40 V toward the intrinsic *Si* (i-*Si*) layer of 1 $\mu$m thickness, where they initiate an electron avalanche, which multiplies the number of electrons in this layer to a readily detectable level. The *p*-doped *Si* of 100 nm thickness forms the charge sheet which ensures that the electric field in i-*Ge* layer is well below an avalanche breakdown field in it, while the field in the *Si* multiplication layer is 3-times greater than the breakdown field in it to provide impact ionization. The structure was grown on highly doped n++ *Si* substrate. The material in i-*Ge* and i-*Si* layers is "pure", i.e. not intentionally doped, that is with concentration of uncontrolled admixtures of ~10$^{15}$ cm$^{-3}$. The device with 25 $\mu$m entrance aperture diameter operated at temperatures T = 100 and 150 K. That SPAD was irradiated by a 10 kHz sequence of 50 ps pulses of laser radiation at wavelengths $\lambda = $ 1.31 or 1.55 μm. The radiation from a laser was sent to the entrance of the detector through a single mode fiber (core diameter of ~10 μm), a calibrated optical attenuator, where the energy in each pulse was reduced to $\leq 0.1\hbar\omega \cong 10^{-20}$ J, and two-lens imaging system. At such conditions DE was measured to be nonzero at both $\lambda = 2\pi c/\omega$ (~4% at $\lambda = $ 1.31 μm and T = 100 K, and ~0.15% at $\lambda = $ 1.55 μm and T = 150 K).

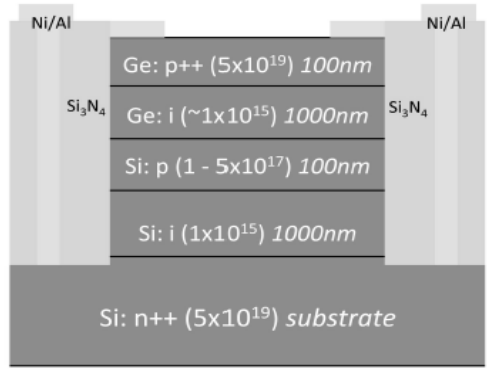

Fig. 1. *Ge*-on-*Si* SPAD structure cross section illustrating two *Ge* layers, two *Si* layers, *Si* substrate, *Ni/Al* contacts, passivation ($Si_3N_4$), doping densities (in brackets), and layer thicknesses (in italics) [5].

The presented description of this SPAD presumes that the upper 100 nm layer of p++*Ge* plays the role of optically transparent electrode and an incident light is absorbed within 1 μm-thick layer of i-*Ge*. The literature search for data on light absorption coefficient ($\alpha$) in *Ge* have shown that $\alpha$ for the incident light in i-*Ge* is of ~(5-6)·$10^3$ cm$^{-1}$ at wavelengths $\lambda \approx 1.3$ μm and ~1.55 μm [11] and of ~(1-2)·$10^3$ cm$^{-1}$ in heavily doped ($\geq 5 \times 10^{19}$ cm$^{-3}$) p-*Ge* at both $\lambda$ [12]. It follows from these data that the p-*Ge* layer is practically transparent ($\alpha l \leq 0.02$), while in i-*Ge* layer at least a half transparent ($\alpha l \leq 0.6$). In light of this at first glance it looks logical that in [5] presumed that PE(s) appear in i-*Ge* layer.

In reality, however, the situation is not so straightforward. The matter is that i-*Ge* is a nondegenerate semiconductor, in which the energy gap, $E_g$, between the valent and conduction zones at T $\approx$ 100K is of ~0.72 eV, and the Fermi energy level is located in the middle of the forbidden zone, $F \cong E_g/2$. Respectively, the temperature induced concentration of free electrons in the conduction band, $N_e$(T=100K), which is expected to be of ~$10^4$ cm$^{-3}$, is negligible compared to $N_e \leq 10^{15}$ cm$^{-3}$ due to uncontrolled admixtures and defects in i-*Ge* which is of ~$10^{15}$ cm$^{-3}$ [13]. It then follows that in the volume of i-Ge layer in [5], $V \approx 1$ μm × (10μm)$^2 \approx 10^{-10}$ cm$^3$, the number of free electrons will be of ~$10^5$. Obviously, it is problematic to detect appearance of a single PE on such background.

The situation is different in a heavily doped p-*Ge* ($N_p \cong (0.5-1)·10^{20}$ cm$^{-3}$) [14]. The Fermi level in such case is shifted to the valent zone and all free electrons are captured by acceptors. As a result $N_e$ tends to zero. In this case appearance of a single additional free electron is obviously an event. It follows from above that the p-*Ge* layer is transparent for incident light ($\alpha l < 0.02$) and has low concentration of free electrons. A material, in which such conditions are upheld, behaves as an optically transparent dielectric.

## 3. The electromagnetic energy

According to [15], the density of electromagnetic (EM) energy, $U_d$, in a dielectric medium is,

$$U_d = \frac{1}{8\pi}\left(\varepsilon E_d^2 + \mu H_d^2\right) \equiv \frac{1}{8\pi}\left(E_d^2 + H_d^2 + 4\pi\chi E_d^2\right), \quad (1)$$

where $E_d$ and $H_d$ are the amplitudes of electric and magnetic fields, $\varepsilon = 1 + 4\pi\chi$ and μ = 1 are the permittivity and permeability, and $\chi$ is the susceptibility. The last term in (1) is the part of EM energy density, which is transferred to movement of the bound electrons (BEs) in a dielectric. It is important to note here that in absence of other losses this energy returns to the radiation field when it leaves a medium. Dividing this energy by the density of BEs number, $N_e$, which are involved into the interaction, one can get the amount of EM energy, which is transferred to one BE, $W_1$. To estimate $W_1$ in the case under consideration we must take into account that the permittivity of *Ge* is $\varepsilon \cong n^2 \cong 17$. This, in particular, means that to sufficient accuracy we can suppose that $U_d \cong U_{in} = I_{in}/c$ where $I_{in}$ is the intensity of incident laser radiation at SPAD entrance in each pulse and $c$ is the velocity of light in vacuum. To estimate $I_{in}$, and respectively $U_d$, we take, for definiteness, the energy in each pulse of $0.1\hbar\omega \cong 1.5 \cdot 10^{-20}$ J and the diameter of irradiated spot at the entrance of device of 10 μm. Then for of 50 ps duration of pulses we get $I_{in} \cong 0.4$ mW/cm$^2$ and $U_d \cong 10^5$ eV/cm$^3$. Respectively, taking into account that the total density of BEs in *Ge* is of $N_e = N_a \times 32 \approx 1.4 \cdot 10^{24}$ cm$^{-3}$, where $N_a$ is the number of *Ge* atoms per cm$^3$, which is of ~$4 \cdot 10^{22}$ cm$^{-3}$, and '32' is the number of electrons in a *Ge* atom, we get $W_1 \cong 10^{-19}$ eV. This energy is obviously much-much less than the forbidden bandgap energy in *Ge*, e.g. $E_g(100K) \cong 0.72$ eV. This actually is true even for radiation pulses with energy $1sh\nu$, $10sh\nu$, $1000sh\nu$, etc.

Then the question raises, namely why and how only one electron never-the-less gets from incident radiation pulse the energy needed to overcome the bandgap barrier in the material.

## 4. The effect of interference

The matter is that an electron driven by an oscillating electric field is the source of a secondary emission. Interference is the only physical phenomenon which is capable to redistribute an averaged energy in a system of many radiation emitters. If we then take into consideration that the incident radiation, which is generated by a laser, is highly coherent throughout the volume of *Ge* layers, $V \approx 1.1 \mu m \times (10 \mu m)^2 \approx 10^{-10}$ cm$^3$, the driven oscillations of electrons in this volume and the reradiated by each of them fields will be coherent too. As such reradiated fields can constructively interfere at some time during irradiation and at some point inside this volume (see e.g. point C in Fig. 2). Taking into account that the total density number of BEs in *Ge* is $N \approx 1.4 \cdot 10^{24}$ cm$^{-3}$, the number of electrons, which are involved into such interaction, is $N \cong N_e \times V \approx 1.4 \cdot 10^{14}$, and potentially the factor of radiation intensity, and EM energy density respectively, enhancement may be up to $N^2 \approx 2 \cdot 10^{28}$. In reality it is much less.

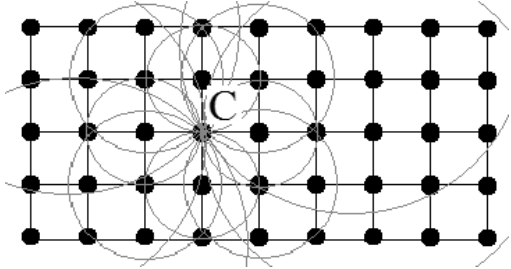

Fig. 2. Schematic sketch of how the constructive interference of EM fields, which are re-emitted by surrounding atoms, may increase radiation intensity at point C in a material lattice. Thin circles represent the wavefronts of fields reradiated by surrounding atoms.

To estimate a potential factor of enhancement we must take into consideration several circumstances. Firstly the nature of secondary emission by electrons is twofold: it may result of an accelerated movement of an electron [16], and of an oscillating dipole which is formed by an electron and a positively charged atomic rest [15,16].

In the first case all ~$1.4 \cdot 10^{24}$ cm$^{-3}$ BEs are sources of re-emission because the electric field $E_i$ of incident radiation moves an electron regardless of where it is located. The strength (amplitude) of an electric field $E_e$, which is re-emitted by a driven with the acceleration $\dot{v}$ electron decreases with distance R from that electron as [16],

$$E_e = \frac{e\dot{v}}{c^2 R} \sin\theta,  \quad (2)$$

where $\theta$ is the angle between the direction of the Hertzian vector and the direction of observation. Correspondent movement is governed by the equation,

$$\dot{v} = \frac{e}{m} E_i e^{-i\omega_0 t}, \qquad (3)$$

where $v$ is the velocity of electron, $e$ and $m$ are its charge and mass and $E_i$ is the amplitude of electric field of incident radiation of frequency $\omega_0$. Respectively we have for the reemitted field amplitude at the distance $R$ from a point charge emitter,

$$E_e = \frac{e^2 E_i}{mc^2 R} \sin\theta, \qquad (4)$$

Let us then evaluate an integral effect of all involved BEs taking into account Eqs (9)&(13). This may be done in the following way. Consider a sphere of radius $R$ with the thickness of wall $\delta$. Its volume is

$$\delta V = \frac{4}{3}\pi\left[(R+\delta)^3 - R^3\right] \cong 4\pi R^2 \delta. \qquad (5)$$

The number of electrons, which are contributing to the field at point C from the distance R inside *Ge* lattice, may be estimated as

$$\delta N_e = N_e \delta V = 128\pi X^2 dX \qquad (6)$$

where $X = R/a_0$ and $dX = \delta/a_0$, and $a_0 \cong 2.8 \cdot 10^{-8}$ cm is the *Ge* lattice constant. Multiplying these by $E_e$ from Eqs (4) and integrating over $dX$ we get the field, which may be induced at point C by the coherent summation of re-emitted by all electrons in *Ge* lattice,

$$E_{e\Sigma} = E_i \frac{64\pi n e^2}{mc^2 a_0}\left(X_m^2 - 1\right) = E_i F_e(X_m), \qquad (7)$$

where $a_0 X_m = R_m$ is the maximum distance from C to re-emitting atom. The magnitude of $|F_e(X_m)|^2$ determines a magnitude of $W_1$ enhancement at point C when re-emission of all electrons in the irradiated volume are coherently summed.

In reality, however, that volume is not coincides with that determined by geometrical sizes of LAS, i.e. $1.1 \times 10^2$ μm$^3$. In particular, because of attenuation of radiation in *Ge*, $R_m$ of order of $1/\alpha \approx 1.7$ μm and respectively $X_m = R_m/a_0 \cong 6 \cdot 10^3$. We then must take into account that, since a PE is expected to appear in p-*Ge* layer, the shape of this volume is the half of sphere of radius $R_m$ and the factor "$\sin\theta$". These two factors may reduce $F_e(X_m)$ by about an order of value. Substituting magnitudes of parameters one would then get $|F_e(X_m)|^2 \cong 10^9$, which is clearly much less than the desirable ~$10^{19}$.

In the second case an atom is considered as a dipole oscillator, the strength (amplitude) of the electric field radiated by which is decaying with $R$ as [15,16],

$$E_d = er(t)\left(\frac{1}{R^3} + i\frac{\omega_0}{cR^2} - \frac{\omega_0^2}{c^2 R}\right)\sin\theta, \qquad (8)$$

where $r(t)$ is the displacement by the electric field of incident radiation of an electron from its equilibrium position at an orbit in an atom. This displacement may be described by the Sellmeyer oscillator equation [17]

$$\ddot{r} + \gamma_i \dot{r} + \omega_g^2 r = \frac{e}{m} E_i e^{-i\omega_0 t}, \qquad (9)$$

where $\gamma_t$ is the coefficient, which characterize a loss of oscillations energy due to inelastic collisions of an electron with surrounding particles and material lattice, and $\omega_g$ is the resonant frequency of an oscillator, magnitude of which is determined by bounding force between the electron and its atomic rest. When the radiation is monochromatic its solution is

$$r_a = \frac{eE_i}{m(\omega_g^2 - \omega_0^2 - i\gamma_i\omega_0)}. \quad (10)$$

It follows from Eq.(10) that, since $\gamma_i \ll \omega_g$, $r_a$ maximizes when $\omega_0 = \omega_g$ and decreases proportionally $1/\omega_g^2$ when $\omega_g \gg \omega_0$. It then follows from Eq.(10) that dipole-kind re-emission by the electrons, which occupy the deeper orbits, may be considered negligible since these have much higher energies of bound with an atomic rest, i.e. much higher $\omega_g$. Consequently, in a *Ge* atom the number of BEs, which are active in the dipole-kind interaction with NIR optical radiation is limited to 4 giving the density of correspondent BEs $N_d = 4N_a \approx 1.8 \cdot 10^{23}$ cm$^{-3}$. Then the number of dipoles, which are contributing to the field at point C from the distance $R = a_0X$ inside *Ge* lattice, may be estimated as

$$\delta N_d = 0.1 N_d \delta V = 1.6\pi X^2 dX. \quad (11)$$

By multiplying these by $E_d$ from Eqs (8)&(10) and integrating over $dX$ we get the field, which may be induced at point C by the coherent summation of re-emitted by the dipoles in *Ge* LAS,

$$F_d(X_m) = \frac{1.6\pi e^2}{mn\gamma_i\omega_g a_0^3}\left[i \ln X_m - \frac{\omega_g a_0 n}{c}(X_m - 1) - i\frac{\omega_g^2 a_0^2 n^2}{2c^2}(X_m^2 - 1)\right]. \quad (12)$$

Among the parameters, magnitude of which determine $|F_d(X_m)|^2$, the most uncertain one is $\gamma_i$. According to [14] its magnitude in semiconductors may vary in the range from ~$10^{10}$ at a room temperature to ~$10^7$ s$^{-1}$ at lower T. Substituting to Eq.(12) magnitudes of $e$, $m$, $n$, $c$, $a_0$, $\omega_g = E_g/\hbar = 10^{15}$ s$^{-1}$ and $X_m \cong 6 \cdot 10^3$, we get

$$|F_d(X_m)|^2 = \left(\frac{3.7 \cdot 10^{18}}{\gamma_i}\right)^2, \quad (13)$$

It follows from Eq.(13) that at $\gamma_i = 10^9$ s$^{-1}$, which a reasonable magnitude for operation temperature of SPAD in [5] (T = 100-150 K), $W_1|F_d(X_m)|^2 \geq 0.7$ eV, i.e. is just the energy sufficient for an electron to overcome the energy barrier of the forbidden zone in *Ge*. An important condition for realization of such enhancement, which is $\omega_0 = \omega_g$, is just the well-known Einstein's condition $\hbar\omega_0 = E_g = \hbar\omega_g$.

## 5. The effect of heavy doping

The specific feature of heavily doped semiconductors, an example of which is p-*Ge* in [5], is that the typical for non-doped materials conventional sharp zone boundaries (dashed lines in Fig.3) are blurred and the "tails" of allowed occupation states penetrate to the forbidden zone resulting in a substantially narrower effective bandgap $E_{ge}$. [18]. The physical reason for this effect is a local fluctuations of the internal electric field in a material due to a generic inhomogeneity of spatial distribution of a dopant at high its concentration [14,18]. The local shift of zone boundaries, which is induced by fluctuations of the internal electric field, is demonstrated in Fig.3*a*) by two solid lines. The shaded areas represent the "tails", which are a result of averaging of the field fluctuations. Fig.3*b*) illustrates schematically the averaged densities of allowed occupation states $N_{c,v}(E)$ for electrons (*c*) and holes (*v*) for pure (dash-dotted lines) and for highly doped (solid lines) material, and the effective bandgap $E_{ge}$.

The problem is, however, that the solid curves in Fig.3*a*) are parallel, which means that a local gap remains unchanged for direct interband transitions (vertical arrow in Fig.3*a*)), and a reduction of the bandgap takes place for indirect transitions (inclined arrow in Fig. 3*a*)) only. In the latter case the efficiency of interband transitions is reduced substantially [14].

In the case of SPAD in operation the situation changes because of applied gating voltage (GV). The homogeneous electric field in p-*Ge* layer, which is induced by that voltage, and which is of ~$5 \cdot 10^4$ V/cm [5], leads the reorganization of zones in a semiconductor in the way, which may be depicted (see Fig.3*c*)) as turning of the picture in Fig.3*a*) [14]. In a result the indirect transition in Fig.3*a*) can become a direct one as in Fig.3*c*) and the gap for this transition is less than $E_g$.

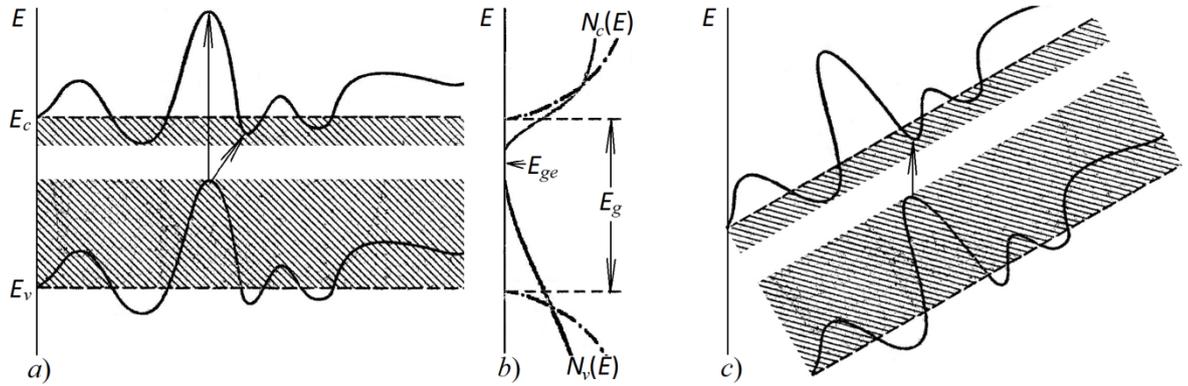

Fig.3. *a*) & *c*) A spatial variation of the conduction and valence band boundaries, $E_{c,v}$, in a pure semiconductor (dashed lines), in a heavily doped semiconductor (solid lines) *a*) without and *c*) with GV applied; vertical an declined arrows illustrate direct and indirect transitions. *b*) Effective densities of allowed occupation states $N_{c,v}(E)$ for electrons (*c*) and holes (*v*) for clear (dash-dotted lines) and for highly doped (solid lines) material.

In particular, in *Ge* at $N_d \geq 10^{20}$ см$^{-3}$ the factor of effective gap narrowing may be up to ~100 [14]. These circumstances allow account for why appearance of a photoelectron under irradiation of a light pulse with sub-photon energy in the SPAD under consideration does not contradict the conservation of energy principle in this interaction.

## 6. Conclusions

Classical macroscopic electrodynamics allows account for the photoelectric effect in a *Ge*-on-*Si* SPAD at a sub-photon energy in incident pulsed laser radiation. The energy of incident laser radiation, which is transferred to a huge number of electrons in *Ge* matrix, can concentrate on only one of these through the effect of the constructive interference of the fields re-emitted by surrounding electrons. The conventional necessary condition for the photoelectric effect in a material, which reads as $\omega_0 = E_g/\hbar$ [7], comes to the model as a resonant condition for the Sellmeyer classical oscillator model. The energy in this interaction is conserved because of a substantial narrowing of the effective bandgap in heavily doped p-*Ge* layer of the SPAD.

**Acknowledgements**

The author would like to thank Professor G. S. Buller for useful discussions.